\documentclass[useAMS,usenatbib]{mn2e}
\usepackage{epsfig,epstopdf,color,ulem,appendix}
\usepackage{bm}
\usepackage{CJK}
\usepackage{color, colortbl}
\definecolor{Gray}{gray}{0.8}
\title{Towards self-consistent modelling of the Sgr A* accretion flow: linking theory and observation}

\begin{document}
\begin{CJK*}{UTF8}{gbsn}
\author[Roberts et al.]{Shawn R. Roberts$^{1}$\thanks{E-mail: srrobert@astro.umass.edu}, Yan-Fei Jiang(姜燕飞)$^{2}$\thanks{Einstein fellow.}, Q. Daniel Wang$^{1}$, Jeremiah P. Ostriker$^{3}$\\
$^{1}$Department of Astronomy, University of Massachusetts, Amherst, MA 01002, USA\\
$^{2}$Kavli Institute for Theoretical Physics, University of California, Santa Barbara, CA 93110, USA\\
$^{3}$Department of Astronomy, Columbia University, 550 W. 120th
Street, New York, NY 10027, USA\\
\vspace{-0.5cm}}
\date{Submitted to MNRAS, July, 2016\vspace{-0.5cm}}

\maketitle

\label{firstpage}
\begin{abstract}
The interplay between supermassive black holes (SMBHs) and their environments is believed to command an essential role in galaxy evolution. The majority of these SMBHs are in the radiative inefficient accretion phase where this interplay remains elusive, but suggestively important, due to few observational constraints.  To remedy this, we directly fit 2D hydrodynamic simulations to \textit{Chandra} observations of Sgr A* with Markov chain Monte Carlo sampling, self-consistently modelling the 2D inflow--outflow solution for the first time.  We find the temperature and density at flow onset are consistent with the origin of the gas in the stellar winds of massive stars in the vicinity of Sgr A*.  We place the first observational constraints on the angular momentum of the gas and estimate the centrifugal radius, $r_\mathrm{c}$ $\approx$ 0.056 $r_\mathrm{b}$ $\approx8\times10^{-3}$ pc, where $r_\mathrm{b}$ is the Bondi radius.  Less than 1\% of the inflowing gas accretes on to the SMBH, the remainder being ejected in a polar outflow.  We decouple the quiescent point-like emission from the spatially extended flow.  We find this point-like emission, accounting for $\sim4$\% of the quiescent flux, is spectrally too steep to be explained by unresolved flares, nor bremsstrahlung, but is likely a combination of a relatively steep synchrotron power law and the high-energy tail of inverse-Compton emission.  With this self-consistent model of the accretion flow structure, we make predictions for the flow dynamics and discuss how future X-ray spectroscopic observations can further our understanding of the Sgr A* accretion flow.
\end{abstract}

\begin{keywords}
accretion, accretion discs - Galaxy: centre - X-rays: general - methods: data analysis
\vspace{-0.5cm}
\end{keywords}

\section{Introduction}

Supermassive black holes (SMBHs) spend the majority of their lives in a low-luminosity/accretion phase ($\sim$90\%; \citealt{Yuan2014}).  A common way to refer to these objects is as low-luminosity active galactic nuclei, or LL-AGN.  The emission from these LL-AGN is very much enigmatic, largely due to the inherent difficulty of separating the various entangled emission components \citep{Wang2013,Li2015}.  Predominantly, there are two unique radiative X-ray phenomena that characterize the emission from LL-AGN: their extremely low, spatially extended quiescent luminosity, and flares that can briefly increase their luminosity by up to a factor of 100 approximately bidaily \citep{Baganoff2001,Yuan2016}. Physically, even though these black holes (BHs) are associated with much lower net accretion rates, they may also be associated with strong mechanical feedback phenomena such as giant radio bubbles \citep{McNamara2016} and collimated outflowing winds.  However, connecting the physical processes around BHs to the observed feedback effects in a self-consistent manner remains a challenge.

As all accreting and outflowing material must flow through the quiescent, spatially extended accretion flow, its emission will be the focus of this study.  For the sake of observational capabilities, this includes both the spatially extended accretion flow and the unresolved point-like emission processes (minus the detected flares) immediately surrounding the SMBH.  Further, we have a perfect candidate to study LL-AGN emission in our own Galaxy, Sgr A*, which already has a wealth of data available through a wide range of wavelength bands \citep{Serabyn1997,Falcke1998,Baganoff2001,Baganoff2003,Hornstein2002,Zhao2003,Marrone2007,Eckart2012,Wang2013,Broderick2016}. 

In X-rays, the Sgr A* quiescent accretion flow is spatially resolved to $\sim 1.4$ arcsec \citep{Baganoff2003}.  This radius, estimated from the analysis of projected X-ray emission in the surrounding field, is roughly consistent with the classical Bondi radius, $r_\mathrm{b}$, enabling the estimate of the rate at which gas is captured by the BH ($\dot{M}_\mathrm{b}\sim10^{-5}$ M$_\odot$ yr$^{-1}$).\footnote{Throughout this paper, the b subscript denotes at the Bondi radius.}  The Bondi radius is an atavism to the original way these objects were studied.  Bondi accretion, from which it derives its name, assumes the accretion flow to be spherically symmetric, with approximately zero angular momentum \citep{Bondi1952}.  In this scenario, any ambient material that enters the Bondi radius, the radius at which the gas thermal energy is equivalent to the gravitational potential energy, is doomed to fall into the BH.  Unless the angular momentum of the gas is very large, the centrifugal radius (classically speaking, the radius at which the gas must transfer angular momentum in order to continue its inward spiral) of the accreting gas will be well within the Bondi radius, making 
this estimate of gas capture rate a reasonable characteristic value for the inflow of gas.  It would be prudent to note that for steady-state accretion, as assumed in this paper, the canonical picture of the centrifugal radius may break down.  As detailed in \cite{Bu2014}, transfer of angular momentum leads to the depression and steepening of the angular momentum profile.  However, regardless of the dynamical nuances within the accretion flow, this value can still be used to characterize the angular momentum at flow onset, that is, the magnitude of angular momentum in captured gas.

Yet, there are several reasons we should tread lightly when considering the angular momentum of the accretion flow.  From a qualitative point of view, the flow does have some apparent flattening \citep{Wang2013}, which in and of itself is suggestive against using Bondi accretion to explain the flow's emission in its entirety.  Indeed, it is quite natural that the accreting gas would have coherent angular momentum, given its apparent origin.  Spectral evidence suggests that the matter accreting on to Sgr A* is shocked stellar wind material emanating from a cluster of O and Wolf--Rayet stars \citep{Wang2013}, a significant fraction of which orbit Sgr A* in a well-constructed stellar disc around the BH \citep{Beloborodov2006}, oriented with an inclination of 127$^\circ(\pm2^\circ)$ and a line-of-node position angle of 99$^\circ(\pm2^\circ$; east from north).  There is also very recent evidence from the Event Horizon Telescope supporting the same orientation for the BH spin axis \citep{Broderick2016}.  However, simulations of the stellar wind dynamics suggest the centrifugal radius occurs well within the Bondi radius \citep{Cuadra2015}, indicating we can still trust, at the very least, this gross estimate of the gas capture rate.

Given the extremely low luminosity of the flow, this capture rate points to the intriguing fact that the radiative efficiency must be extremely low, $\ll0.01$, loosely dubbing the m\'{e}lange of remedying models radiatively inefficient accretion flows (RIAF).  While the detailed properties of these flows may differ significantly, in order to be considered a RIAF flow, the model of BH accretion must merely satisfy the faint requirement.  The most common models under this umbrella are characterized by accretion-comparable outflows and/or the advection of energy into the BH (for a recent review, see \citealt{Yuan2014}).  By advecting energy into the BH, or driving it away via an outflow, the luminosity of the accretion flow itself is naturally lessened immensely.  Further, these models have been shown to explain the basic features of the quiescent emission from Sgr A*, including radiative efficiency, low-resolution spectrum \citep{Yuan2014}, and multiwavelength spectral energy distribution (SED; \citealt{Yuan2003}).  

The physical reality of the Galactic Centre makes one model in particular quite attractive, the rotating, radiating inflow--outflow solution, or RRIOS \citep{Narayan2012,Yuan2012b,Li2013,Yuan2015,OstrikerInPrep}.  This is primarily for two reasons.  First, the apparent origin of the gas is suggestive that angular momentum is important.  Second, as found with Faraday rotation measurements, the amount of gas that accretes on to the BH, $\dot{M}_\mathrm{BH}$, is $\leq$0.01 $\dot{M}_\mathrm{b}$ \citep{Marrone2007}.  This is also the conclusion reached by \cite{Melia2001}, who calculated the total inverse-Compton emission of the inner disc assuming a Bondi accretion profile and found that it significantly overpredicted the observed X-ray emission.  Li et al. (2013) demonstrated that for RRIOS-like flows the bulk of the inflowing gas flows out again after reaching roughly the centrifugal radius, resulting in a small rate of net accretion on to the BH.  These models can take a range of angular momenta, spanning from pure Bondi accretion to anything that leaves the centrifugal radius reasonably within the Bondi radius, with no direct observational constraint as of yet.  The determining of this angular momentum, however, has dire implications for the strength and distribution of any outflowing mechanical feedback (Ostriker et al., in prep).

Perhaps the most in-depth X-ray observational study of the quiescent emission from Sgr A* thus far has been done by \cite{Wang2013}.  This is thanks to a recent wealth of data from the \textit{Chandra} X-ray Visionaries Program, which provides over 3 Ms of Sgr A* observations.    In that paper, the authors study the BH from a purely spectral perspective and are able to place some constraint on the X-ray emission.  Unfortunately, that work was unable to place any constraints on the angular momentum of the gas.  However, they show that the spectrum suggests a fitted radial density profile that is consistent with a strong outflow, nearly balancing the inflow, using an approximate 1D analytical RIAF model.  The authors were even able to place limits on the deconvolution between quiescent point-like and extended emission, showing that unresolved residual point-like emission (with detected flares removed) can only account for $\leq$20\% of the quiescent emission.  While we expect undetected flares to contribute relatively little to this point-like quiescent emission from an extrapolation of flare fluences, emission processes very near the event horizon are the most uncertain, and therefore need not be the case.

Thus, in order to understand some of these very local processes near the SMBH, we should keep in mind the flares, since a truly unified model would be able to explain both the flare emission and the quiescent point-like emission.  Further, understanding of one may help to illuminate the conditions of the other. In the study discussed above, the authors also show that the cumulative spectrum of the flare emission is observed to be a powerlaw with index $\sim2.6$ \citep{Wang2013}.  While the production mechanism of flare emission is still poorly understood (e.g., \citealt{Yuan2016}), the variability time-scale makes it clear that they originate in localized regions very near the SMBH.  This creates many challenges from a theoretical standpoint, as the physics near a SMBH are an extreme in the Universe, leaving their physical origin up for debate.  

However, some compelling work has recently been done by \cite{Ball2016} to understand the flare emission.  After previous studies have suggested the importance of thermodynamically decoupling electrons from ions at low radii \citep{Yuan2003}, these authors created the first relatively large-scale general relativistic magnetohydrodynamic (GRMHD) simulations to include a subgrid prescription for modelling the non-thermal electrons.  Operating under the assumption that the accretion flow is rotationally supported within their simulation boundaries has allowed them to simulate, and generally match, the detected multi-wavelength flaring properties of Sgr A*.  They show that the flares can be explained by particle acceleration in highly magnetized regions through magnetic reconnection.  Further, they provide the first roughly unified model of an LL-AGN at low radii.  Thus, by extension, they also provide a general framework for understanding the quiescent point-like emission.  In general, as we move to lower radii, the magnetic field becomes increasingly important.  When considering models, we should allow for the two different populations of electrons: a larger thermal population and a much smaller non-thermal population in a highly turbulent environment.  Thus, possibly important emission mechanisms include the full spectrum: bremsstrahlung \citep{Wang2013}, inverse-Compton \citep{Liu2001,Yuan2003}, and synchrotron emission \citep{Liu2001,Yuan2003}.

Clearly, despite all of the observations, the details of the dynamical dance of the shocked stellar wind gas with the BH is still held under relatively few constraints, leaving alternative models to continue marauding the theoretical landscape (e.g., \citealt{Yusef2016}).  This means we need to undertake serious effort to observationally fetter the flow properties.  However, genuine observational constraint cannot be obtained in this case without legitimate modelling, both spatial and spectral.  Since this means highly non-linear dynamical modelling, it requires stepping out of the analytical realm and into simulation.  This is needed not only to understand the multidimensional flow structure required to self-consistently model the inflow and outflow simultaneously, but also the magnetic field structure, which is likely to be an important factor for understanding the emission that originates very near the BH.  Simulations of these kinds of accretion flows have been done in at least some capacity for several years now \citep{Narayan2012,Yuan2012,Li2013}, but taking these simulations and connecting them to observation in a self-consistent manner is a challenge and has yet to be done.
  
By doing just that, albeit with slightly simplified 2D hydrodynamic simulations, we herein attempt to remedy as many of the uncertainties as reasonably possible and provide a path forward for deepening our understanding of the Galactic Centre.  We compare images of the combined quiescent {\it Chandra} data (the same data set as analysed in \citealt{Wang2013}) directly to simulations of BH accretion via the development of a suite of Markov chain Monte Carlo (MCMC) tools designed for this purpose. By breaking the quiescent emission into several bands, we can use as much information in the data as possible, utilizing both spectral and spatial power simultaneously.  This serves to break degeneracy between key interesting flow parameters such as temperature, density, angular momentum, and inclination angle, providing some of the first self-consistent constraints on not only the accretion flow structure, but also the decomposition of spatially extended accretion structure from point-like emission.  Such an analysis will provide a more legitimate test of the RRIOS solution, as well as the broader class of RIAF models, and lend significant guidance to the way that we model the processes surrounding LL-AGN in the centre of galaxies.

\section{Methods}
\label{s:methods}

The \textit{Chandra} X-ray Visionary Project (XVP) to observe the Milky Way’s SMBH, Sgr A* from 2012 February 6 to October 29 resulted in approximately 3 Ms of exquisite data, opening up a completely new regime of insight into BH accretion.  \textit{Chandra} and the XVP program have offered us an excellent opportunity to observe directly the dynamics of hot gas around a BH, and, with proper modelling, hopefully elucidate the inner workings of the enigmatic class of objects known as LL-AGN.  Combining this observational data with 2D RRIOS simulations \citep{OstrikerInPrep}, we attempt to constrain directly the accretion structure via the power of Bayesian MCMC fitting.

\subsection{Data Preparation}

\begin{figure}
\centerline{\epsfig{figure=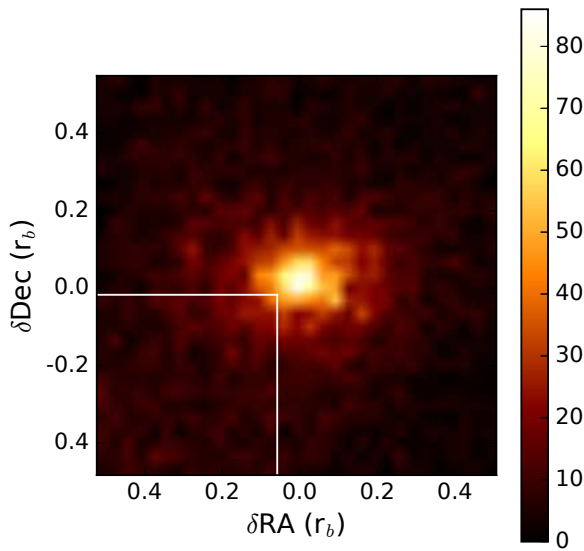,width=0.5\textwidth,angle=0}}
\caption{
Counts image of Sgr A* in the 1.-9. keV band taken with a 2775 578 s exposure using \textit{Chandra}. Colour bar represents total counts in a pixel.}
\label{f:obsfull}
\end{figure}

For a more detailed description of the data reduction and quiescent X-ray image generation, we recommend the reader to \cite{Wang2013}.  However, in short, the data are reduced via standard \textsc{ciao} processing routines (version 4.5; Calibration Database version 4.5.6).  Since the differences between individual observation pointings are all within 14 arcsec, the merged data are treated as a single observation.  \cite{Wang2013} found no apparent calibration issues.  Flares are removed from the quiescent data through detection with the `\textsc{bayesian blocks}' routine \citep{Neilsen2013}, leading to a total quiescent exposure time of 2.78 Ms.  The observed quiescent image over the entire spectral band (1-9 keV) can be seen in Fig. ~\ref{f:obsfull}.  The southeast corner of the image is excluded in the fit due to significant emission from an unmodelled feature in the region (highlighted by white lines in Fig. 1; see also Wang et al. 2013).   The region used for fitting extends to a radius of approximately 0.5 $r_\mathrm{b}$.  Since the source is on-axis, the full width at half-maximum (FWHM) of the instrument is comparable to the pixel size.  To utilize additional spatial information from the dithering of observations, we construct the so-called `super-resolution' image with a pixel size of 0.123 arcsec on a side (Wang et al. 2013).  The full band image is then split into three bands, to place some spectral constraints on the fit and reduce degeneracy.  The bands (1--4,4--5.5, and 5.5--9 keV) were chosen such that the counting statistics are roughly similar in each image.

\subsection{Hydrodynamic Simulations}

We solve the same set of hydrodynamic equations in 2D spherical polar coordinate as in \cite{Li2013} using the new radiation MHD code {\sc athena++} (Stone et al., 2016, in preparation). {\sc athena++} is an extension of the multi-dimensional MHD code {\sc athena} with the new capability of curvilinear coordinates and logarithmic grid.  This allows us to cover a large dynamic range, as was possible with the older \textsc{zeus} code used in \cite{Li2013}, but still solve the hydrodynamic equations with the higher order Godunov method. Since we are only simulating in 2D, explicit kinetic viscosity (fixed to be $10^{-3}c_{\mathrm{s},\infty}r_\mathrm{b}$, where $r_\mathrm{b}$ is the Bondi radius and c$_{s,\infty}$ is the ambient sound speed, as in \cite{Li2013}) is used to mimic the angular momentum transfer caused by magneto-rotational instability.  This viscosity corresponds to a dimensionless angular momentum transport efficiency, $\alpha$, $\sim0.01$ near the centrifugal radius. The simulation setup is also similar to \cite{Li2013} with radial range covering 10$^{-3}$ $r_\mathrm{b}$, approx. 400 Schwarzschild radii, to 10 r$_b$ with 1024 logarithmically spaced grid cells.  In the polar coordinate, $\phi$ varies from $0.6^{\circ}$ to $179.4^{\circ}$ to avoid singularity along the pole and it is divided into 256 uniform grid cells. A reflecting boundary condition is used along the poles. This polar grid is the primary difference from the simulations done by \cite{Li2013}, where symmetry was also imposed with respect to the equatorial plane.

 The temperature, density, and radial velocity of the simulation domain is initialized to the Bondi profile with an ambient temperature set to be $1.16\times 10^7$ K. The central BH mass is assumed to be $4.1\times 10^6$ M$_{\odot}$.  These parameters imply $r_\mathrm{b}\approx3.7$ arcsec, which we will assume to be fixed as in \cite{Wang2013}.  Bremsstrahlung cooling is included as in \cite{Li2013}.  There are two free parameters to consider when setting up the simulation: the density at $r_\mathrm{b}$ and the centrifugal radius, $r_\mathrm{c}$.  The density at $r_\mathrm{b}$ determines the Bondi accretion rate and the cooling time-scale compared with the local dynamic time-scale while $r_\mathrm{c}$ sets the angular momentum of the inflowing gas, as well as the initial angular momentum of gas in the simulation domain (excluding the poles). Our simulations confirm the basic conclusion of \cite{Li2013}.  When the Bondi accretion rate is below $\sim 0.01$ Eddington accretion rate, we obtain the hot solution with net accretion rate at the centre of the simulation domain smaller than $1\%$ of the Bondi accretion rate. We also find significant outflow along the polar direction. However, the main difference between our simulations and results shown in \cite{Li2013} is that we do not find any outflow along the equatorial plane. We confirm that if we impose symmetry along the equatorial plane, the equatorial outflow shows up, which suggests that this is an artefact of the imposed symmetry. In the hot solution regime, properties of the solution, such as total emission, density and temperature profiles, scale with the density at $r_\mathrm{b}$ for a fixed $r_\mathrm{c}$, which is proportional to the Bondi accretion rate. When running the simulations, we choose a density scaling at $r_\mathrm{b}$ such that Bondi accretion rate is $10^{-3}$ of the Eddington accretion rate.  This ensures that we are in the hot solution domain, which makes the solutions scalable in density due to self-similarity.  We have run simulations spanning a range of centrifugal radii to explore how the solution changes with $r_\mathrm{c}$, from 0.01 $r_\mathrm{b}$ to 0.2 $r_\mathrm{b}$.  It is from these simulations that all models of the accretion flow are derived.  A more complete description of the 2D RRIOS hydrodynamic solutions will be presented in another paper \citep{OstrikerInPrep}.

\subsection{Fitting Procedure}

We use the power of Bayesian inference through MCMC sampling to fit the simulated accretion flows detailed above to the \textit{Chandra} observations.  One of the beauties of Bayesian inference is that by introducing the idea of subjective probability, it provides a framework to incorporate prior information.  In the case of the physical sciences, this prior information often represents things such as boundary conditions, or past fitting results.  Further, by treating parameters themselves as random variables, we directly sample a model's reality given the data, or the posterior, providing us with fully described confidence boundaries for each model parameter.  This is an important point, because rather than traditional approaches which can only seek to exclude, we are testing the affirmation of a model and its parameters.

In its most simplistic terms, Bayesian inference can be formalized as
\begin{equation}
P(\boldsymbol{\theta}|\boldsymbol{D})= \frac{P(\boldsymbol{D}|\boldsymbol{\theta})P(\boldsymbol{\theta})}{P(\boldsymbol{D})}\propto P(\boldsymbol{D}|\boldsymbol{\theta})P(\boldsymbol{\theta})
\end{equation}
where $\boldsymbol{\theta}$ represents our model parameters, $P(\boldsymbol{\theta})$ represents our prior belief in the model components, $P(\boldsymbol{D}|\boldsymbol{\theta})$ is the probability of the data given the model (or the likelihood function), $P(\boldsymbol{D})$ is the probability of the data (also called the evidence), and $P(\boldsymbol{\theta}|\boldsymbol{D})$ is the probability of the model given the data (the posterior).  The evidence term, which requires integrating out the model over all parameter space, places significant constraint on what can be done analytically. Thankfully, it can be neglected when discussing computational approximation, such as that calculated with MCMC techniques.

In this paper we assume all the data are generated from the underlying model with Poisson probability:
\begin{equation}
P(D_k|\boldsymbol{\theta})=\frac{\lambda(\boldsymbol{\theta})^{D_k}e^{-\lambda(\boldsymbol{\theta})}}{D_k!}
\end{equation}
where $k$ is the $k$th pixel, $D_k$ is the photon count in pixel $k$, and $\lambda$ is the expected number of counts for a given pixel, which is calculated as the exposure multiplied by the flux as determined below.  By sampling directly from the posterior we are able to then obtain confidence intervals for our parameters.  In this analysis, we use the Metropolis--Hastings algorithm to sample the posterior probability distribution for the parameter set.  This is a MCMC method that relies on rejection sampling to obtain a sequence of random samples drawn from the posterior, and can thus be used to approximate the posterior probability distribution.

\subsection{Modelling}

\begin{figure}
\centerline{\epsfig{figure=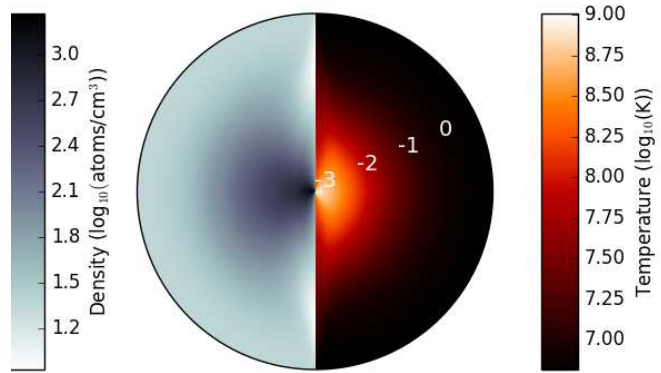,width=0.5\textwidth,angle=0}}
\caption{
Time averaged density (left) and temperature (right) distributions for the best-fitting solution.}
\label{f:td_dists}
\end{figure}

\begin{figure}
\centerline{\epsfig{figure=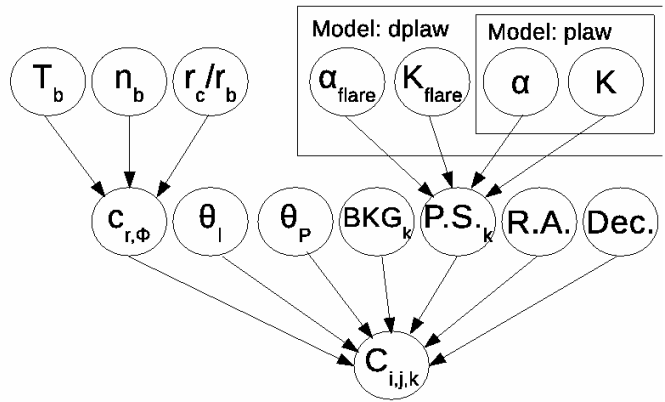,width=0.5\textwidth,angle=0}}
\caption{
The hierarchical Bayesian network.  Bondi temperature and density combine with the centrifugal radius to form a volume normalized count rate for each grid cell, $c_{r,\phi}$.  Flare power-law index and normalization combined with an unknown power-law index and normalization to form a point source emission in each band.  These hyperparameters combine with the background (BKG$_k$) in each band, the positioning on the detector (R.A. and Dec.), the inclination angle ($\theta_\mathrm{I}$), and the projection angle ($\theta_\mathrm{P}$) to form total counts in each pixel, $C_{i,j}$.}
\label{f:BayesNet}
\end{figure}

The full list of fitted parameters and their relationship to each other is pictorially described in Fig.~\ref{f:BayesNet} and a summary of all the models to be compared is shown in Table~\ref{t:msum}.  The full list of priors used in this paper can be found in Table~\ref{t:priors}.\footnote{In the following notation, $i$ and $j$ represent the pixel coordinates and $k$ represents the $k$th band image.}  In order to fit the simulated accretion flows to the \textit{Chandra} data, we generate synthetic images from an interpolation of the hydrodynamic simulations and employ a hierarchical Bayesian fit, with three levels of parameters.  Assuming a metallicity as that of the local Sgr A* complex, $\mu\sim0.76$, we only require three parameters to fully characterize the accretion flow's emission since they are approximately self similar: temperature scaling -- $T_\mathrm{b}$, density scaling -- $n_\mathrm{b}$, and centrifugal radius -- $r_\mathrm{c}$/$r_\mathrm{b}$\footnote{For these simulations, parameterizing values in terms of the Bondi radius is simply used for convenience.}.  In this context, $r_\mathrm{c}$ represents the magnitude of the gas angular momentum at capture.  The spatial density and temperature distribution of the accretion flow is generated by interpolating time averaged hydrodynamic simulation data (see Fig. \ref{f:td_dists}) at different $r_\mathrm{c}$/$r_\mathrm{b}$, in this case, through 0.04, 0.08, and 0.2, allowing us to sample a continuous distribution in $r_\mathrm{c}$.\footnote{We have also fit the images interpolating through $r_\mathrm{c}$/$r_\mathrm{b}$ = 0.02, 0.04, and 0.08 with no change in results.}  It should be noted that while this is the case for almost all accretion flow models in this paper, we will have a separate model (model $pBondi$, for pseudo-Bondi) where the angular momentum is fixed to the lowest of our simulations in order to directly test the realism of the Bondi solution for Sgr A*.  Then, to build a 3D model of the flow, we need only introduce its positioning and orientation information.  This includes the right ascension (RA), declination (Dec.), inclination angle ($\theta_\mathrm{I}$), and position angle ($\theta_\mathrm{P}$, east from north).

Once we have a 3D model, we need to calculate its emission in order to compare to an observed image.  Volume normalized emission measure as a function of temperature is calculated within \textsc{xspec} using the model \textsc{tbabs*vapec*dustscat}, as used in the spectral fit to the integrated quiescent emission \citep{Wang2013}.  Both absorption and dust scattering effects are taken into account.  In all of the modelling, we assume the absorption column, as estimated from the powerlaw model fitting to the accumulated flare spectrum, is 13.76$\times$10$^{22}$ cm$^{-2}$ (Wang et al. 2013).  After mapping the \textsc{xspec} calculations to the simulation temperature table, and scaling it with the density table, we have the volume-normalized counts for each radius and $\phi$.  Finally, for generating a synthetic image of the accretion flow we integrate along the line of sight at the centre of each pixel out to $r_\mathrm{b}$, unless the pixel contains the origin.  If the pixel contains the origin, we integrate from the inner boundary of the simulation out to $r_\mathrm{b}$.

While the accretion flow accounts for the bulk of the emission in the \textit{Chandra} images, we still need to account for other sources as best we can.  This includes both a spatially smooth background component(BKG$_k$) and a point-like component centred on the BH (P.S.$_k$).  This background accounts for all foreground and background contributions (e.g., including faint stellar and extragalactic sources, as well as diffuse hot gas), which are assumed to be smoothly distributed on scales of a few arcseconds.  Since this smooth background contributes relatively little to the overall flux in the images, a spectral decomposition and modelling of these two is beyond the scope of this paper. Therefore, each $k$th band background component is allowed to roam free with respect to the others.  A characterization of the quiescent P.S.$_k$, however, is of great interest.  Recall, this includes anything within the inner boundary of our simulation, $\sim400r_\mathrm{s}$.  Thus, a model test, comparing several parametrizations of this enigmatic component is required.  In general, with the exception of the $pBondi$ model, different model names refer to a different parametrization of this point-like component.

Since the X-ray emission most proximal the BH remains very uncertain, there are many ways in which we could consider parametrizing the point source contribution.  To get a baseline of what might be the best possible fit, particularly for other parameters, model \textit{free} places no constraint on the relationship between individual P.S.$_k$ components, allowing them to roam free, similar to BKG$_k$.  However, we would like to physically model this point-like emission.  To first order, we might expect this contribution to just be dominated by one component, e.g. unresolved flare emission, inverse-Compton emission, or thermal bremsstrahlung.  Thus we could consider it to be well fit by a single power-law component (models \textit{plaw} and \textit{plaw-wp}).  In this case, we have only two free parameters to characterize the P.S. emission: the photon index, $\alpha$, and a normalization, K.  No prior constraint is placed on this power-law index.

Alternatively, we may consider a scenario where we have multiple competing emission components.  In this case, we assume the flare emission in X-ray has an approximately fixed mean spectral shape, as obtained in \cite{Wang2013}.  Thus, the flare emission has a strong prior on the power law index, $\alpha$ equal to 2.6 $\pm$ 0.4 (90\% confidence; \citealt{Wang2013}).  We assume that any additional point source emission that cannot be attributed to the simulation can be approximately parametrized as a power-law.  This potentially unknown power-law component is left with no prior constraint as to the index.  In this scenario, these two power-law components then combine to form the point source normalization in each band, P.S.$_k$ (model \textit{dplaw}).  

The normalized power-law emission for a given power-law index, $\alpha$, is calculated within \textsc{xspec} using the model \textsc{tbabs*pow*dustscat}.  Again, both absorption and dust scattering effects are taken into account.  Assuming no interloping point sources, P.S.$_k$ and BKG$_k$ combine with the projected, integrated accretion flow emission to create the total pixel counts, $C_{i,j,k}$.  The final step in creating a synthetic image that can be compared to the observed image is to convolve $C_{i,j,k}$ with the \textit{Chandra} Advanced CCD Imaging Spectrometer (ACIS) point spread function (PSF), which is assumed to be described by a Gaussian of FWHM=0.5 arcsec.  \footnote{Since the effective frequencies of the bands are quite similar due to the spectral shape, this should be a sufficient approximation.  Indeed, MARK simulations show only a few percent difference between the energy enclosed in the inner arcsec between the low and high band. See Fig. 4.6 http://cxc.harvard.edu/proposer/POG/html/chap4.html.  This does, however, neglect the Lorentz contribution to the PSF, which causes a significant spread for approximately 15\% of collected photons at this energy.  Thus, we do not expect it to differentially affect the different bands appreciably; however, we may overestimate the background emission.} 

\begin{table}
  \caption{Summary of the models to be compared.  Here pBondi stands for `pseudo-Bondi', plaw for a power-law P.S.$_k$ parametrization, and wp for `with prior', referring to the prior on the flow orientation.}
 \centering
 \label{t:msum}
  \begin{tabular}{ccc} 
Model name & P.S. & Additional comments \\ \hline
free & Independent &  \\
plaw & Single power-law &  \\
plaw-wp & Single power-law & Prior on flow orientation\\
 & & set by stellar disc\\
dplaw & Double power-law & Prior on one power-law slope\\
 & & set by the flare emission slope\\
 pBondi-wp & Single power-law & $r_\mathrm{c}$ is fixed at the lowest \\
 & &  of our sims (0.02 $r_\mathrm{b}$); prior\\
 & & on flow orientation set by\\
 & & stellar disc\\
\end{tabular}
\end{table}

\begin{table}
  \caption{Summary of the priors.  A uniform distribution is represented as U(lower,upper) and a normal distribution is represented as N(mean,sigma).}
 \centering
 \label{t:priors}
  \begin{tabular}{ccc} 
Parameter & Model & Prior \\ \hline
$T_\mathrm{b}$ & All & U(0,$\inf$) \\
$n_\mathrm{b}$ & All & U(0,$\inf$) \\
$r_\mathrm{c}$/$r_\mathrm{b}$ & pBondi-* & 0.02 fixed\\
 & All others & U[0.04,0.2]\\
$\theta_\mathrm{I}$ & *-wp & N(127,2)\\
 & All others & U[90,180]\\
$\theta_\mathrm{P}$ & *-wp & N(99,2)\\
 & All others & U[0,180]\\
BKG$_k$ & All & U[0,$\inf$)\\
P.S.$_k$ & free & U[0,$\inf$)\\
$\alpha$ & *plaw*, dplaw & U[1,10]\\
$K$ & *plaw*, dplaw & U[0,$\inf$)\\
$\alpha_\mathrm{flare}$ & dplaw & N(2.6,0.12)\\
$K_\mathrm{flare}$ & dplaw & U[0,$\inf$)\\
\end{tabular}
\end{table}

\subsection{Numerical Caveats}

The \textsc{apec} implementation within \textsc{xspec} restricts the temperature of the plasma to $\leq68$ keV.  Since the temperatures in the simulation tables span a large range, reaching temperatures significantly higher than this, we need to estimate beyond the allowed \textsc{xspec} temperature.  As the emission from such a plasma is dominated by free-free emission, we can estimate its emissivity using the following equation:
\begin{equation}
\epsilon^\mathrm{ff}_\nu \propto T^{-1/2}\mathrm{exp}\left(-\frac{h\nu}{kT}\right)
\end{equation}
Further, since the spectral shape is approximately constant with increasing temperature in the energy range covered by $Chandra$, we can straightforwardly extrapolate the emission at temperatures $>$68 keV from the absorbed, volume-normalized emission at 68 keV.  Lastly, it should be noted that any error introduced in this term will be absorbed into the central point source contribution, as these temperatures only occur very close to the BH.  However, the density does not increase quickly enough with decreasing radius and the volume occupied by this high-temperature gas is relatively miniscule, making its overall contribution to the flux in the image quite small, $<1\%$.

\section{Results}
\label{s:results}

First, we will determine the best model to scrutinize, as well as assess the goodness of fit.  Ideally, this would be done by a full Bayes factor calculation, which is the ratio of the posterior odds of one model to another (or rather, which model better explains the data, $D$):
\begin{equation}
\mathrm{BF} = \frac{P(M_1|\boldsymbol{D})}{P(M_2|\boldsymbol{D})}
\end{equation}
This is the odds of one model versus the other given the data.  If we do not consider any prior model favouritism, BF reduces to the ratio of model likelihood.  Unfortunately, this ratio is actually quite difficult to calculate, as it requires integrating over all of parameter space for each model.  Thus, since our model parameters are well behaved (more or less normally distributed), we will take advantage of an approximation, the Bayesian information criterion, between models:
\begin{equation}
\Delta \mathrm{BIC}= -2\mathrm{log}\left[\frac{\mathrm{max}_{M_1}P(\boldsymbol{D}|\boldsymbol{\theta}_1)}{\mathrm{max}_{M_2}P(\boldsymbol{D}|\boldsymbol{\theta}_2)}\right]-(p_2-p_1)\mathrm{log}(S)
\end{equation}
where $\boldsymbol{\theta}$ is the maximum likelihood parameter vector for the respective model, $p$ is the number of parameters for the model, and $S$ is the number of data points.  This statistic is shown to approach the Bayes factor for large sample size \citep{Schwarz1978}, before including model priors.  A guide to numerical interpretation of this statistic is shown in Table~\ref{t:dbic} \citep{Kass1995}.

\begin{table}
  \caption{Numerical interpretation of $\Delta$BIC.}
 \centering
 \label{t:dbic}
  \begin{tabular}{cc} 
$\Delta$BIC & Strength of Evidence Against \\ \hline
0-2 & Barely mentionable \\
2-6 & Positive\\
6-10 & Strong\\
$>$10 & Very strong\\
\end{tabular}
\end{table}

A comparison of models through $\Delta$BIC is shown in Table~\ref{t:mcomp}.  We see that the $plaw$ model optimizes the amount of information gained.  Unfortunately, for this model the inclination angle is not strongly constrained, as long as the flow is at least moderately edge-on.  Although the goodness of fit is slightly worse when we add a prior on the flow's inclination and projection angle ($plaw$-$wp$), the $\Delta$BIC is so small ($\sim1.7$) as to not even be mentionable.  Therefore, since we cannot realistically discriminate between these two models, and we heavily favour the $plaw$-$wp$ model a priori, due to our favourability of the stellar wind gas origin, we will consider this model the best model.  This allows us to somewhat more tightly constrain other parameters.  Other models, including the $dplaw$ and $pBondi$ are heavily disfavoured.  The $dplaw$ model is disfavoured because adding extra degrees of freedom does not result in an appreciably better fit, suggesting the point source can be reasonably characterized by a single power-law within the limits of the data, and the $pBondi$ model because fixing the angular momentum to our lowest allowed value leads to a much poorer fit in general.  We discuss the causes and implications of the poor fit of the $pBondi$ model in Section \ref{sec:ang}.

\begin{table}
  \caption{$\Delta$BIC between models, relative to the best-fitting model.}
 \centering
 \label{t:mcomp}
  \begin{tabular}{ccc} 
Base model & With prior & Without prior \\ \hline
free & 6.4 & 4.7 \\
pBondi & 46.4 & N/A \\
plaw & 1.7 & 0.0 \\
dplaw & N/A & 12 \\
\end{tabular}
\end{table}

The results of the fitting process for our best-fitting model, $plaw$-$wp$, are listed in Table~\ref{t:pars} along with the 90\% confidence bounds.  The flow enters the Bondi radius at a temperature ($\sim1.3\mathrm{e}7$ K) and density ($\sim100$ cm$^{-3}$), great enough to be detectable in the X-ray band with such a long exposure provided by the XVP program with \textit{Chandra}.  The gas also has considerable angular momentum, leaving $r_\mathrm{s}\ll r_\mathrm{c}\sim0.056r_\mathrm{b}$ $\approx8\times10^{-3}$ pc.  This density and angular momentum implies a mass inflow rate at $r_\mathrm{b}$, $\dot{M}_\mathrm{in,b}$ of $2.4\times10^{-3}$ $\dot{M}_\mathrm{Ed}$ and a mass accretion rate, $\dot{M}_\mathrm{acc}$, of $\leq10^{-2}\dot{M}_\mathrm{in,b}$ There is a non-negligible, steep residual point-like component to the emission.  This point-like component has a specific luminosity of log$_{10}(\nu L_\nu)\sim31.96$ erg s$^{-1}$ at 5 keV, and is responsible for 4.2 (2.3,7.0)\% of the observed emission within 1.5 arcsec in the 1-9 keV band with \textit{Chandra}.

\begin{table}
  \caption{Best-fitting and 90\% confidence intervals of free parameters marginalized over all others for the model \textit{plaw-wp}.  Model implied quantities are distinguished by the grey rows.}
 \centering
 \label{t:pars}
  \begin{tabular}{cc } 
$\theta_\mathrm{I}$ ($\deg$) & 126.4 (122.6,130.4) \\
$\theta_\mathrm{P}$ ($\deg$) & 99.1 (96.3,101.8) \\
$r_\mathrm{c}/r_\mathrm{b}$ & 0.056 (0.048,0.066) \\
$T_\mathrm{b}$ (K) & 1.28e7 (1.19e7,1.42e7)  \\
$n_\mathrm{b}$ (cm$^{-3}$) & 101.6 (91.4,111.1) \\
BKG1 (counts pixel$^{-1}$) & 0.21 (0.01,0.45) \\
BKG2 (counts pixel$^{-1}$) & 0.41 (0.17,0.65) \\
BKG3 (counts pixel$^{-1}$) & 0.48 (0.30,0.69) \\
$\alpha$ & 4.8 (3.5,7.5) \\
log$_{10}$($K$) (erg s$^{-1}$ at 5 keV) & 31.96 (31.32,32.18) \\
\rowcolor{Gray} $\Rightarrow\dot{M}_\mathrm{in,b}/\dot{M}_\mathrm{Ed}$ & $\sim 2.4\times10^{-3}$\\
\rowcolor{Gray} $\Rightarrow\dot{M}_\mathrm{acc}/\dot{M}_\mathrm{b}$ & $\leq10^{-2} $ \\
\end{tabular}
\end{table}

Looking at the full distribution of each parameter and the parameter--parameter confidence bounds (supplementary materials online) suggests very little degeneracy, except for those which are quite natural: for example, the anticorrelation between temperature and density.  This anticorrelation is very much expected, due to their emissivity proportionality, $\propto n^2 T^{1/2}$ in the relevant temperature range.  Further, the two parameters are well constrained for the first time in a self-consistent fashion, with fairly narrow uncertainty ranges.  With the exception of the power-law index and normalization, all of the parameter PDFs are more or less normally distributed.  This suggests the information provided by the observations herein is sufficient to fully characterize the quiescent accretion flow.

\begin{figure}
\centerline{\epsfig{figure=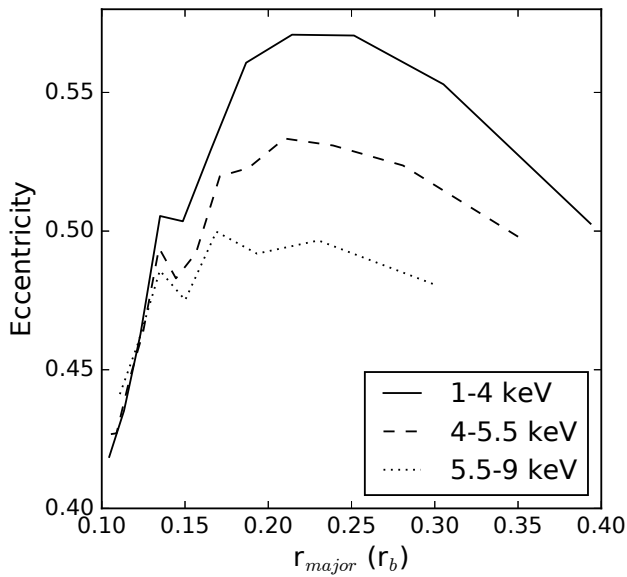,width=0.5\textwidth,angle=0}}
\caption{
Emission eccentricity as a function of major axis radius.  These profiles correspond to the top panel of images in Fig.~\ref{f:images}, the theoretical emission smoothed by the \textit{Chandra} PSF.
}
\label{f:eccentricity}
\end{figure}

\begin{figure*}
\centerline{\epsfig{figure=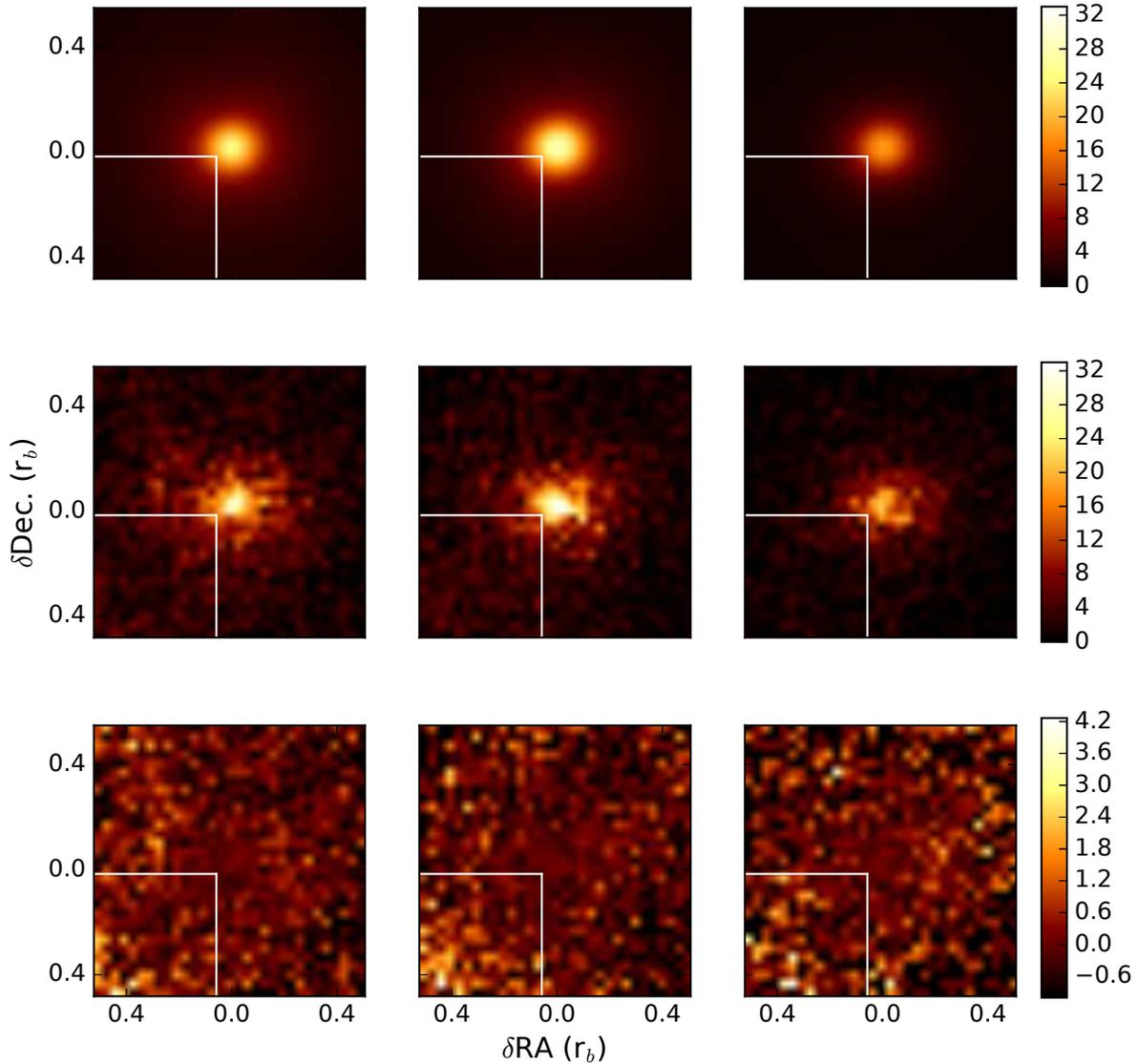,width=1.0\textwidth,angle=0}}
\caption{
From top to bottom: the theoretical image, observed image, and residual image ((observed$-$theoretical)/theoretical).  These are shown for all three bands, which are from left to right: 1-4, 4-5.5, and 5.5-9 keV.
}
\label{f:images}
\end{figure*}

The theoretical images generated from these best-fitting values are shown in Fig.~\ref{f:images} and compared to observed images.  As previously noted, but never quantified, the emission is considerably flattened.  We plot the eccentricity in each observation band as a function of major axis radius in Fig.~\ref{f:eccentricity}.  At low radii, where the emission is dominated by the point-like emission, the eccentricity is very low.  The eccentricity increases steeply up to $\sim$0.56, 0.53, and 0.49 at 0.2 $r_\mathrm{b}$ for the 1-4, 4-5.5, and 5.5-9 keV band, respectively, where the emission is predominantly from the extended accretion flow. As we then move to larger radii, the eccentricity begins to decrease as the background emission becomes increasing important. 

We see that there are no apparent residual effects, particularly at lower radii.  However, there can be contributions from unmodelled structure at large radii ($\geq1$ arcsec): for example, the spur of emission to the north-east of Sgr A*.  This is echoed in the goodness of fit estimate (Appendix~\ref{app:goodness}).  In the inner arcsecond, where the predominant source of error results from time averaging the simulations and not modelling the turbulent structure, the statistical consistency is at a level of 2.6\%.  When including radii out to $\sim 2.8$ arcsec, where there are many more error terms to consider, such as unmodelled extended structure and discrete X-ray sources, the statistical consistency drops.  However, we have found the fitted parameters insensitive to the radius of the fitted region within the observational images shown in Fig.~\ref{f:images}.  Thus, all uncertainties considered, we consider this fit to be not only reasonable, but quite good.

\section{Comparison with Previous Works and Implications}

Now, we need to discuss these numerical results in the context of previous modelling and observational efforts.  While prior observations have not lent us a great deal of constraint, they have provided some important information for modelling of the quiescent accretion flow.  This includes constraints on the mass inflow at different radii, the radiative efficiency of the accretion flow, and some 1D considerations about the structure of that accretion flow.  From a theoretical perspective, simulations have separately made significant strides in attempting to model the physics very near the SMBH, as well from its apparent origin in stellar winds.  With our results here, we are able to significantly build upon these past observational constraints and help to lend guidance to future modelling efforts.

\subsection{Temperature/density and the gas origin}

At flow onset, i.e., $r_\mathrm{b}$, the best-fitting temperature and density are consistent with previous estimates based on spectral analysis of the \textit{projected} X-ray emission as observed by \textit{Chandra} \citep{Baganoff2003}.  They are also consistent with the simulation results of \cite{Cuadra2015}.  These authors attempt to model the flow onset by simulating the stellar wind dynamics of the surrounding stellar cluster.  By estimating stellar mass loss, they show that the density at $r_\mathrm{b}$ is $\sim$100 cm$^{-3}$.  Similarly, their predicted temperature is $\sim$ 1$\times$ 10$^{7}$ K at $r_\mathrm{b}$.  This can easily be understood in the context of shocked stellar winds.  Our best-fitting value temperature implies a shock velocity of $\sim$1000 km s$^{-1}$, which is reasonably characteristic of stellar winds.  For example, the wind velocities range from approximately 600 to 2500 km s$^{-1}$ in the simulations of \cite{Cuadra2015}.  Thus, our result is consistent with an origin of the mass flow in shocked stellar winds.

The temperature and density radial profiles of the accretion flow inferred from our best-fitting simulation are also consistent with previous X-ray spectroscopic estimates.  \cite{Wang2013} approximated the profiles in a RIAF model as power-laws ($n\propto r^{-3/2+s}$ and $T\propto r^{-\theta}$).  Their spectral analysis gives the best fit $\gamma=2s/\theta = 1.9(1.4, 2.4)$. If $\theta = 1$, via the virial theorem, then $s\sim1$, indicating a very flat density profile of the flow, or an outflow mass-loss rate that nearly balances the inflow \citep{Wang2013}.  They assumed this parametrization characterizes the flow over a wide range of radii, between $r_\mathrm{in}$ and $r_\mathrm{out}$ of $\sim10^2$ $r_\mathrm{s}$ and $\sim10^5$ $r_\mathrm{s}\approx0.25r_\mathrm{b}$, respectively.  Even though the simulated profiles we used are not strictly power-laws (e.g., Fig.~\ref{f:dens_comp}), they are roughly in agreement with their conclusion.  We find that the density profile is ever so slightly steeper ($s\sim0.93$), and temperature profile is mildly flatter ($\theta\sim0.77$).  Together, these are consistent with the relation above, within uncertainty.

\subsection{The need for angular momentum and $r_\mathrm{c}$}\label{sec:ang}

\begin{figure}
\centerline{\epsfig{figure=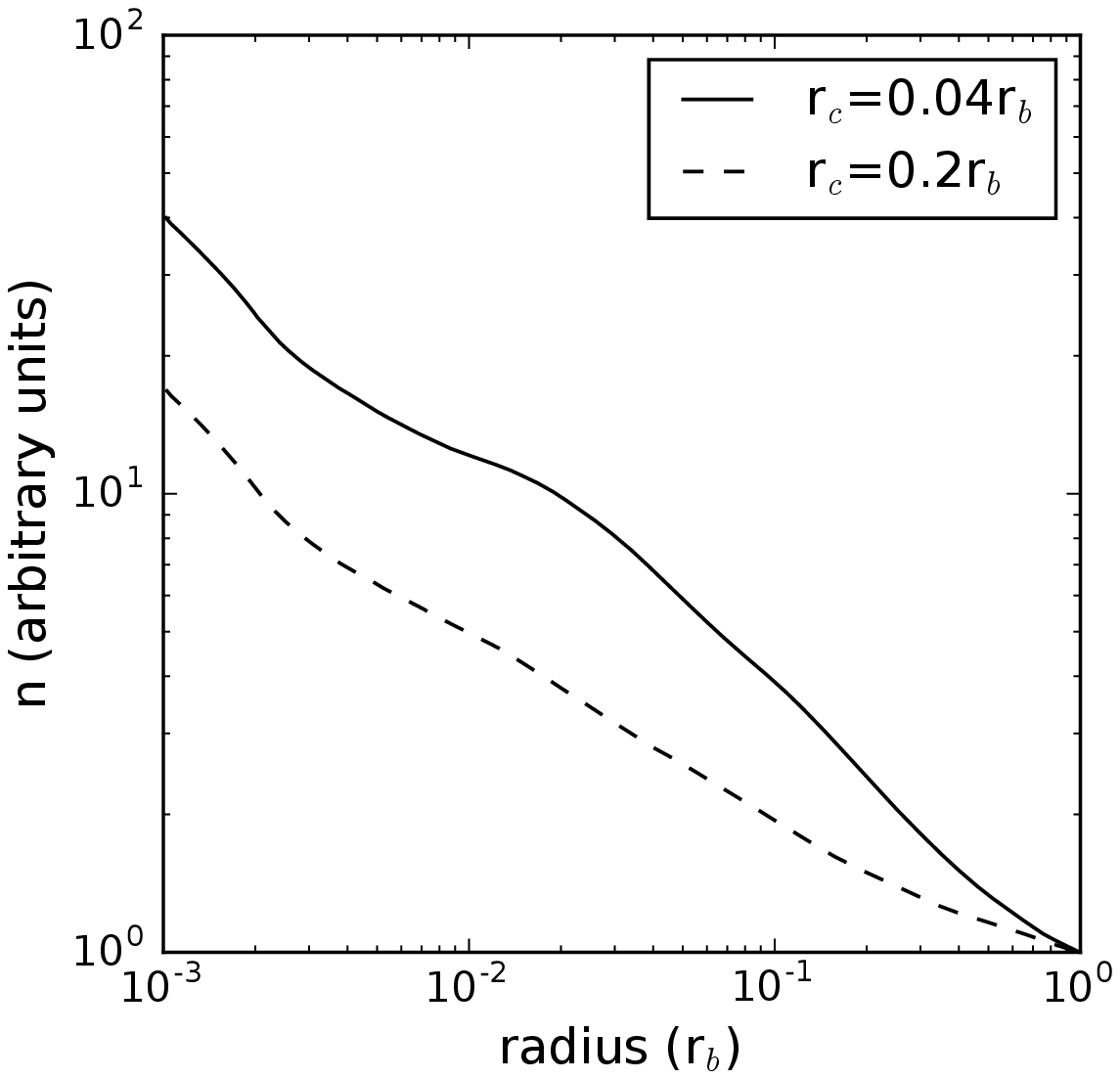,width=0.5\textwidth,angle=0}}
\caption{
Azimuthally averaged density profile for two different simulations, normalized to 1 at $r_\mathrm{b}$.  A lower angular momentum leads to a steeper density profile.}
\label{f:dens_comp}
\end{figure}

Since spherical Bondi accretion is still occasionally invoked when trying to understand the accretion flow on to Sgr A* (e.g., \citealt{Rozanska2015}), or LL-AGN in general, we need to test and understand exactly why this is a poor assumption.  Compared to our best fit, a lower angular momentum accretion flow leads to a steeper density profile. This can easily be seen in Fig.~\ref{f:dens_comp}, which shows the azimuthally averaged density profile for two different angular momentum solutions, and is in agreement with previous results \citep{Bu2013}.  There are two primary forces that support gravity in an accretion flow, the gas pressure gradient and the centrifugal force.  As the centrifugal force decreases, this necessitates a larger gas pressure gradient, and thus a steeper density profile.

\begin{figure}
\centerline{\epsfig{figure=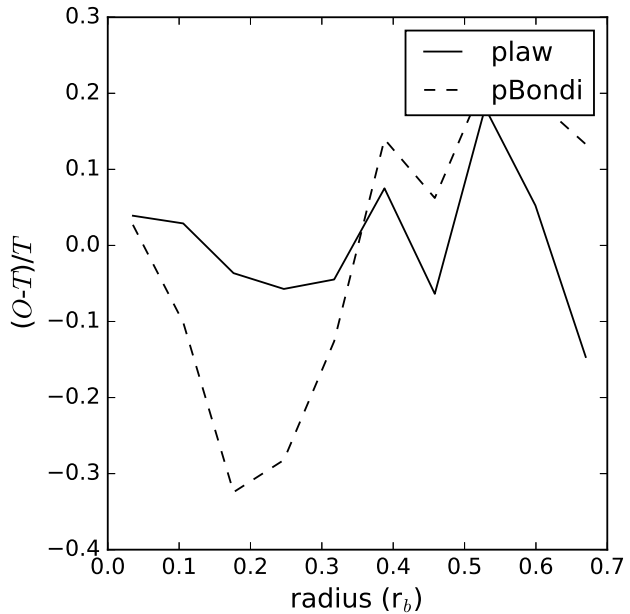,width=0.5\textwidth,angle=0}}
\caption{
Mean residual ((observed$-$theoretical)/theoretical) plotted as a function of radius for two models in the 1--4 keV energy band.  The $pBondi$ model has difficulty modelling the emission at intermediate and large radii simultaneously.}
\label{f:images_pbondi}
\end{figure}

This density profile change manifests itself in the best-fitting solution to other parameters.  In order to compensate, the $pBondi$ model has a decreased Bondi capture density of $\sim$40 cm$^{-3}$, approximately a factor of 2 below the best-fitting model.  The density change largely comes from the central pixels in the medium- and high-energy bands.  For fixed Bondi density, the steep rise of density with radius leads to too much emission very near the BH.  In an attempt to offset the decrease in density, the temperature is increased slightly to 1.6$\times$ 10$^{7}$ K to help model the flux at larger radii.  This in turn makes the emission of the accretion flow harder, which pushes the PS power-law to a spectral index of ~9($\pm$0.5).  However, these changes fail to fully compensate for the low angular momentum; the rapid increase in density of this simulation cannot realistically model the emission at both small and large radii simultaneously.  This is most prominently seen in Fig.~\ref{f:images_pbondi}, which shows the mean residual as a function of radius, comparing the $pBondi$ model and the $plaw$ model.  We can see that there is strong residual structure at intermediate to large radii.  Further, these best-fitting temperature and density values are near, if not beyond in the case of temperature, the limits of what is possible given the previous simulations of \cite{Cuadra2015}.  As the angular momentum decreases further, it is likely that these values would become irreconcilable with other results.

At a best-fitting $r_\mathrm{c}$=0.056 $r_\mathrm{b}$, the centrifugal radius sits at $\approx$ 20000 $r_\mathrm{s}$.  This value has hitherto only been estimated once theoretically, but never observationally.  Simulating the onset of the accretion flow by modelling the stellar wind dynamics around Sgr A*, \cite{Cuadra2008} predicted $r_\mathrm{c}$=5000$r_\mathrm{s}$.  This seeming discrepancy is likely a numerical result, for multiple reasons, all stemming from the fact that they made this estimate using the very inner region of the flow.  First, their simulations were evolved with the smoothed particle hydrodynamics (SPH) code, \textsc{Gadget}-2, which has known issues modelling angular momentum in the depths of a potential well \citep{Keres2009}. Further, by the author's own admission, they suspect the number of SPH particles could be too small to realistically estimate the angular momentum in the inner region of the flow that was used to do so.  Lastly, as we know from \cite{Bu2014}, for a steady accretion scenario the centrifugal radius does not exist in any material sense, but merely represents a characteristic radius for the magnitude of the gas angular momentum at the \textit{outer} boundary.  In reality, the angular momentum profile depresses and steepens relative to the classical picture due to the transfer of angular momentum.  As such, any estimate of $r_\mathrm{c}$ made with the inner flow will naturally underestimate $r_\mathrm{c}$.  This can be seen in \cite{Cuadra2008}, where they show that after initializing the simulation, their average angular momentum of the inner region spikes to a level that is loosely consistent with our determination of $r_\mathrm{c}$, but then dwindles to one fourth of that value as simulation time progresses.  We will also point out that the density profile of a flow with such a low angular momentum is unable to model the spatial distribution of emission.  The centrifugal radius of the $pBondi$ model is slightly larger than the result predicted by \cite{Cuadra2008}, and as we show above, even a flow with this value is well outside the bounds of reality.

Since the multidimensional structure of the flow is largely encapsulated by the determination of the gas angular momentum, as parametrized by $r_\mathrm{c}$, the results detailed here are of prime importance.  Being the first observational constraint on $r_\mathrm{c}$, our result is an important one for simulations that wish to study the outflow but cannot realize the flow from its origins.  Unfortunately, there is not yet a 3D simulation large enough to model the flow through such a dynamical range, simultaneously self-consistently producing the flow and generating the outflow. Simulations that do realize the flow from its origins will be important for verifying our results, especially as they continue to become more realistic.

\subsection{The central point source}

There have been many works that attempt to constrain the unresolved point-like residuals.  Studying the surface brightness profile, \cite{Shcherbakov2010} estimate a point-like contribution of $\sim10\%$.  An extrapolation of flare fluences below the detection limit \citep{Neilsen2013} and a statistical analysis of the X-ray flux distribution \citep{Neilsen2015} also find a similar residual point-like contribution.  \cite{Wang2013} do two separate analyses trying to place constraints on the unresolved point source emission.  By comparing the radial intensity profiles of a flare image with the quiescent image they place a limit on the PS emission to $<$20\% of the total X-ray flux within $\leq$1.5 arcsec \citep{Wang2013}.  In a separate analysis, they spectrally decompose the point-like emission from the extended emission.  That spectral decomposition suggests the unresolved point source emission, assumed to be due to bremsstrahlung, contributes 16(5, 23)\% to the total flux.  Our best-fitting model, placing the fraction of unresolved point-like emission in this region at 4.2(2.3, 7.0)\%, is well below the upper limit constraint from spatial decomposition and loosely consistent with the result from spectral decomposition.

While the total flux is roughly in agreement, the spectral shape is quite the contrast.  When assuming this unresolved emission is a power-law, we find that it is characterized by a spectral index, $\alpha = 4.8(3.5, 7.5)$, grossly steeper than the assumed bremsstrahlung spectrum in the spectral decomposition of \cite{Wang2013}.  Thus, we are able to rule out bremsstrahlung as an important emission mechanism in the inner 100 $r_\mathrm{s}$.  Recall, we noted that  $<$1\% of the emission in the images originated between radii of 100 and 1000 $r_\mathrm{s}$ since the density does not rise quickly enough to compensate for the low volume and rising temperature, further suggesting that bremsstrahlung should not be important for the emission so near the SMBH.

The discrepancy between our findings and those of \cite{Wang2013} can be understood quite naturally when looking across all of the differences.  Not only is their PS spectrum significantly harder than the PS deconvolution in our analysis, but it is also a much more significant contributor to their flux.  This results in the PS emission pushing their flow emission to a softer spectrum.  It does so by flattening the density profile and steepening the temperature profile, both of which put more emission in a cooler flow component than the analysis presented here.  This discrepancy illustrates the problem with modelling spectrally alone, which entangles emission components with too many degrees of freedom, where we have to make too many assumptions to make progress.  To break the degeneracy, we need to have some constraint on at least one component solely, point-like or extended flow, as is done here with the outskirts of the flow.

Another possibility is that the point-like emission is unresolved flare emission.  Yet, here too, lies a discrepancy.  The cumulative flare spectrum exhibits a spectral index of $\alpha = 2.6 \pm 0.2$, as seen in \cite{Wang2013}.  They also show that there is no evidence of a significantly changing spectral index with flare strength.  Therefore, unless the properties of unresolved flares are substantially different from resolved ones, this spectral index is well beyond the bounds of certainty placed by our analysis here.  We deem this unlikely, due to the apparent universal nature of X-ray flare emission around a BH, not just for Sgr A*, but for the general population of LL-AGN \citep{Li2015}.

In reality, there are likely two primary contributors to the quiescent X-ray point-like emission, synchrotron and inverse-Compton scattering, which can be understood in the context of both Ball et al. (2016) and Yuan et al. (2003).  These theoretical works set the stage for completing the picture of quiescent emission near Sgr A*.  By decoupling electrons and allowing them to become non-thermal, Ball et al. (2016) show that the flares could naturally be due to trapped particles in magnetic flux tubes, which are accelerated through reconnection.  But we can also see from that work that these particles are only a small fraction of the overall electron population.  The rest of the electrons exist in a very hot, either thermal or quasi-thermal, turbulent environment.

This interpretation is qualitatively in agreement with the calculations of Yuan et al. (2003). They show that the multiwavelength quiescent SED can be explained by electrons in a quasi-thermal distribution.  The bulk of the electrons are thermal, emitting strong synchrotron emission in the radio bands.  Some of this thermal synchrotron emission is inverse-Compton upscattered into the ultraviolet (UV), with a high-energy exponential tail extending into the X-rays. Also, in their model, approximately $\leq$1.5\% of the electrons must be accelerated into a synchrotron power-law tail in order to match the quiescent infrared (IR) emission.  These electrons, which exist outside of the strong magnetic flux tubes associated with the flares, are in an approximate steady state, with synchrotron cooling times typically greater than the advection time-scale.  Further, Yuan et al. (2003) show that the power-law index of the synchrotron emission must be greater than or equal to 3.5.  This scenario is consistent with the multiwavelength SED spanning from radio through IR and to X-ray, including more recent estimates of the mean IR flux \citep{Schodel2011}.

The power-law slope found in our work, at 4.8 (3.5, 7.5), is in agreement with this predicted upper limit of 3.5 from \cite{Yuan2003}, if only slightly more steepened.  Since we would expect the very steep thermal inverse-Compton emission to be detected predominantly in our 1--4 keV band, it is reasonable that our power-law would be steepened slightly.  However, since the residuals in Fig.~\ref{f:images} do not show any drastic residuals at the origin that would result from a significant deviation of the point-like emission from a power-law, we believe it is reasonable to conclude the emission is primarily due to synchrotron and the inverse-Compton upscattering of this non-thermal emission, with slight contamination from thermal inverse-Compton emission in the 1--4 keV band.

However, their predicted flux is in direct conflict with the results detailed herein.  Based on their calculations, we expect the thermal inverse-Compton emission to contribute anywhere from a few tenths of a percent to a percent of the X-ray emission, and, is naturally quite steep due to its thermal origin.  A power-law index of 3.5 places the synchrotron contribution to X-ray emission at roughly a few percent.  Some of this non-thermal synchrotron emission is also inverse-Compton scattered to X-ray energies, contributing $\sim10\%$ to the quiescent emission with approximately the same slope as the synchrotron emission.  Thus, we expect the total combined synchrotron and inverse-Compton emission to contribute approximately 10--20 percent to the quiescent Sgr A* emission from the model of \cite{Yuan2003}, approximately four times greater than what we observe.

However, there are several model differences that we believe would serve to negate this issue.  Most notably is the density profile, $\rho\propto r^{-3/2+s}$, assumed in \cite{Yuan2003} for their 1D analytical RIAF solution is much steeper, $s\sim0.27$, than we find in X-ray studies, $s\sim1$, from this work and that of \cite{Wang2013}.  The change in density slope requires the population of ultrarelativistic particles to lessen significantly, decreasing both the synchrotron emission and the inverse-Compton flux.  However, making this change in their model creates some other outstanding issues.  Specifically, it would lead to an underprediction of the observed sub-mm emission.  Yet, these issues we believe to be amenable, given a treatment of the multidimensional structure of the accretion flow.  For one, we know from \cite{Ball2016} that the magnetic field strength is much greater in the polar outflow region than assumed to be throughout the flow in \cite{Yuan2003}.  Secondly, the assumption of a Maxwellian distribution of electrons at each radius likely leads to further mistreatment of the outflow region.  We will explore the full implications of these model differences via an update to the SED model of \cite{Yuan2003} in a separate paper, as well as explore the physical nature of the steep synchrotron.  However, we believe the model changes discussed above qualitatively provide a path for reconciliation.

\section{Model Predictions and Future Work}

Understanding the low radiative efficiency of Sgr A* is of central importance to learning about the processes surrounding LL-AGN and how their feedback affects the circumnuclear environment.  In recent years, the general picture of how the accretion flow evolves has begun to emerge.  As gas spews forth from Sgr A*'s large, circumnavigating O and Wolf-Rayet stars in the form of stellar winds, it collides with other stellar winds, shocking to temperatures that greatly ionize the gas, causing it to emit in X-rays.  Without the angular momentum to resist the gravitational lure of Sgr A*, it is captured by the BH, and begins falling deeper into the potential well.  Whether that gas circularized or not was unclear. However, we show that it indeed does have (and requires) coherent angular momentum, circularizing well within the capture radius, but still quite distant from the BH.  As angular momentum is transported, the gas turbulently dances closer to the BH.  Some of this gas will accrete on to the BH.  Yet, most of it will be driven away in a large collimated polar outflow, to what distance is unclear.  With a general framework now in place, we can begin to look in more depth at the implications of this observationally constrained accretion flow and how we may further test the model.

\subsection{Flow dynamics}

\begin{figure}
\centerline{\epsfig{figure=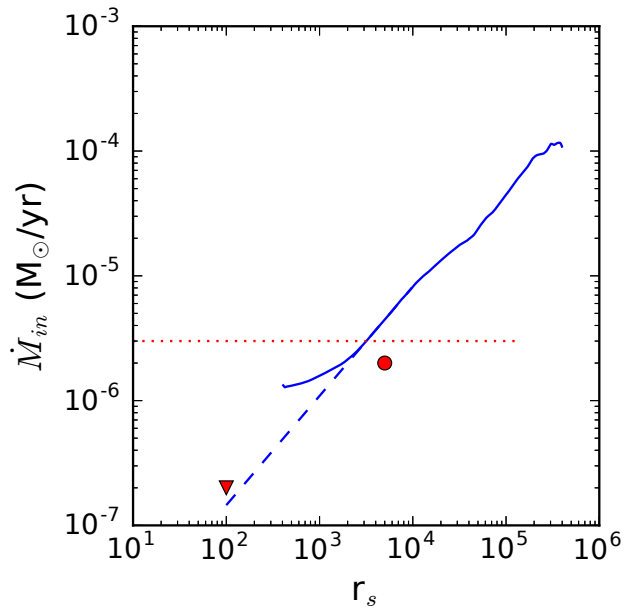,width=0.5\textwidth,angle=0}}
\caption{
The solid line gives the inflow mass flux, temporally and azimuthally averaged, as a function of radius.  The dashed line provides an extrapolation to lower radii.  The red dotted line is the estimate made by \protect\cite{Baganoff2003}, based on the cumulative emission within 2$\times10^5$ $r_\mathrm{s}$, assuming Bondi accretion.  The circle is the estimate made with the simulations of \protect\cite{Cuadra2008}.  The upper limit placed by \protect\cite{Marrone2007} is the upside down triangle.
}
\label{f:accrate}
\end{figure}

The azimuthally averaged gross mass inflow rate is shown in blue in Fig. \ref{f:accrate}.  The curve is nicely consistent with other estimates of the mass accretion at different radii, both theoretical and observational.  The simulated estimate by \cite{Cuadra2008} lies very near our curve, shown by the red circle.  This curve is also roughly consistent with the results of \cite{Baganoff2003}, which is shown by the red dotted line.  In that work, the authors assume accretion is Bondi-like, and estimate the rate based on the cumulative spectrum within 1.5 arcsec.  Compared to the RRIOS model, which has a relatively flat density profile and corresponding steep mass inflow profile, their Bondi assumption places much more gas at low radii and naturally must underestimate the accretion rate at $r_\mathrm{b}$ in order to compensate.

Another important constraint on the accretion flow is that applied to low radii by \cite{Marrone2007} based on radio data.  They place an upper limit of $\sim2\times10^{-7}$ M$_\odot$ yr$^{-1}$ at 100 $r_\mathrm{s}$ based on estimates from radio polarization.  Unfortunately, this is inside the inner boundary of our simulations. We directly estimate the inflow rate to be $\sim10^{-6}$ M$_\odot$ yr$^{-1}$ at $\sim10^3$ $r_\mathrm{s}$ from the simulations.  Extrapolating our curve to 100 $r_\mathrm{s}$ places the inflow rate at $\sim1-2\times10^{-7}$ M$_\odot$ yr$^{-1}$ (with the inferred net accretion rate less than or equal to this value), just inside the limit placed by \cite{Marrone2007}.  Note, the flattening of the inflow rate at low radii in the simulation is an artificial feature, due to its proximity to the inner boundary.  Thus, we have neglected it in the extrapolation.

  We predict a mass inflow rate of $\sim10^{-4}$ M$_\odot$ yr$^{-1}$ $\approx2.4\times10^{-3}\dot{M}_\mathrm{Edd}$ at $r_\mathrm{b}\approx4\times10^5r_\mathrm{s}$.  This rate is well within the limits placed by estimates of stellar mass loss in the vicinity of Sgr A*.  There are $\sim30$ stars that have important mass loss rates \citep{Paumard2006,Cuadra2015}, with individual mass-loss rates in the range $5\times10^{-6}-10^{-4}$ M$_\odot$ yr$^{-1}$ \citep{Martins2007, Cuadra2008}.  This inflow rate at $r_\mathrm{b}$ is also approximately an order of magnitude below that required for the hot accretion flow solution \citep{Li2013}.

The mass outflow rate roughly follows the mass inflow rate as a function of radius, leaving the two in approximate balance, and creating an approximately constant net mass accretion rate.  This means that at all radii, the \textit{net} mass accretion is extremely low.  Approximately 1\% of the material that is accreted at $r_\mathrm{b}$ makes it to radii of $10^3$ $r_\mathrm{s}$, with the rest being driven out in the polar outflow.  This outflow has a large opening angle, defined to be the angle in $\phi$ that has a positive time-averaged radial velocity, of $\sim130$-$140^\circ$.  The density weighted velocity of the outflow is $\sim350$ km s$^{-1}$.  This velocity is undoubtedly an underestimate.  With the inclusion of magnetic fields and the self consistent generation of feedback, we expect this velocity to increase.  Thus, until this type of fit is done with a 3D MHD simulation, it is unclear how much kinetic energy is stored in the polar outflow.  Further, we are still unable to determine where this energy will be deposited and therefore exactly how much feedback it represents.

\subsection{Observational predictions}

Hopefully, with the launch of \textit{Athena}, made even more necessary after the breakup of \textit{ASTRO-H}, we will have an immensely powerful new tool to probe the hot universe.  Unfortunately, the spatial resolution of the \textit{Athena} instrument is only $\sim5$ arcsec, comparable to $r_\mathrm{b}$ for Sgr A*.  However, the effective area and spectral resolution (2.5 eV; goal of 1.5 eV) are much improved.  Thus, by analysing line profiles, an observation of the Sgr A* complex with \textit{Athena} will be able to provide some very good dynamical information about the flow.  Further, at this spectral resolution, different lines can be used to probe different regions of the flow.  This is particularly true if the goal of 1.5 eV is met, which would give enough resolution to have diagnostic power for lines spanning from 2 to 7 keV.

\begin{figure}
\centerline{\epsfig{figure=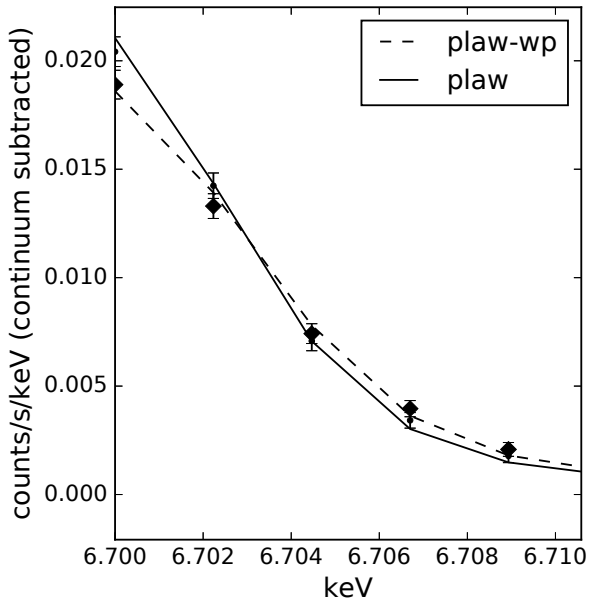,width=0.5\textwidth,angle=0}}
\caption{
Simulated line emission (continuum subtracted) for the Fe \textsc{XXV} K$\alpha$ resonance line at 6.7 keV observed with \textit{Athena} for 500 ks.  Bin size is $\sim2.25$ eV.  Diamonds correspond to simulated bins for the $plaw$-$wp$ model, and dots to the $plaw$ model.}
\label{f:lprof}
\end{figure}

For example, let us consider the strong He-like Fe \textsc{XXV} K$\alpha$ resonance line at 6.7 keV \citep{Wang2013}.  A spectral resolution of 2.5 eV is equivalent to $\sim100$ km s$^{-1}$ at 6.7 keV.  This is incredibly good resolution, considering for our model, the gas that emits in the Fe \textsc{XXV} line has velocities up to $\sim500-2000$ km s$^{-1}$.  We have simulated this line's emission for a 500 ks observation of Sgr A* (Fig. \ref{f:lprof}) for both the $plaw$ and $plaw$-$wp$ models (ignoring bulk turbulent motions, which are only $\sim$15\% of the rotational velocity).  Note, we have only displayed the high-energy side of the line, as the low-energy side will be contaminated by other Fe \textsc{XXV} transitions.  However, since the plasma is optically thin, it will be symmetric about 6.7 keV.  We see that the two models can easily be distinguished from each other using the line profile for this depth of observation.  Thus, line profiles as observed with \textit{Athena} can be used to independently constrain the inclination angle of the accretion flow.  Not only does this illustrate a way to incorporate more information and greatly constrain the Sgr A* accretion flow, but also highlights the immense diagnostic power of \textit{Athena}.

Even though \textit{Athena} promises to be a great leap forward, it will still leave a fair amount to be desired for those who want to do spatially resolved spectroscopy.  However, such an instrument is not outside the realm of reality.  The proposed \textit{X-ray Surveyor} would be the proverbial holy grail of X-ray astrophysics.  With spectroscopic resolution comparable to that of \textit{Athena}, spatial resolution similar to that of \textit{Chandra}, and $\sim50$ times better effective area as that of \textit{Chandra}, we would be able to map individual lines within $r_\mathrm{b}$.  Spatially modelling many of these lines simultaneously would provide incredible constraint on the structure of the accretion flow, potentially allowing us to reconstruct the 3-D inflow and outflow structure.

\subsection{Faraday rotation measurements}

If the type of fitting in this paper is done with a 3D MHD simulation, then we can utilize other observations to constrain the accretion flow, most notably, Faraday rotation measures.  This measure is the integral of the product of the density and parallel magnetic field along the line of sight, $\propto\int n_\mathrm{e}\boldsymbol{B}\cdot\mathrm{d}\boldsymbol{l}$.  This measure has already been successfully used to place limits on the accretion rate very near the BH.  Assuming some basic structure and energy equipartition, Marrone et al. (2007) place a lower and upper limit of 2e-9 M$_\odot$ yr$^{-1}$ and 2e-7 M$_\odot$ yr$^{-1}$ near the BH, respectively.  Others, such as Li et al (2015), have attempted to make calculations based on some toy models, in an attempt to understand the accretion and outflow processes, but have met with contradiction between their estimate of the flow's inclination angle and that of the stellar disc.

With the new RRIOS simulations, we now have an understanding of the general distribution of material around the BH.  In principle, we could use the rotation measure as a constraint with the current 2D simulation that we have used, assuming some limiting cases of the magnetic field geometry.  While we are not willing to make such assumptions at this time, these magnetic fields will be generated self-consistently in a 3D MHD simulation, giving us more information to leverage in our quest to understand the low accretion phase and its mechanical feedback.  However, at this time we leave such work for a future paper.

\subsection{Further numerical considerations}

We have discussed many reasons why it is important to do the type of fit presented here with a 3D MHD simulation, including Faraday rotation estimates and the self-consistent generation of feedback and outflow velocity.  We should, however, point out an additional numerical concern.  Our current simulations, in using an artificial viscosity, may not be modelling the transfer of angular momentum as accurately as we would like.  While previous work suggests the inclusion of magnetic field has little effect on global properties such as the density profile \citep{Begelman2012,Yuan2012}, any change in the transfer of angular momentum would have strong consequences for the radial density profile, which is the primary constraint on the angular momentum through the centrifugal radius.  Therefore, we believe that the global parameters, particularly the angular momentum of the flow, which is tightly constrained here, could change.  However, care has been taken to make the simulated flows as realistic as possible.  But, in any case, detailed 3D MHD simulations should be carried out, starting with the best-fitting parameters, to check various consistencies and to address the role of magnetic fields.

\section{Summary and conclusions}

While self-consistently connecting the outflowing gas to the surrounding circumnuclear area remains an outstanding challenge, significant strides have been made in recent years modelling the physics surrounding low-luminosity BHs, in particular Sgr A*.  Since it is numerically infeasible to simulate from very near the BH through the accretion flow to the origin of the feeding material and its subsequent outflow deposition area, many orders of magnitude in resolution, the community has resorted to modelling in specific spatial domains.  \cite{Yuan2015} and \cite{Ball2016} have made great strides recently in simulating the physics very close to the BH, and from the other end, Cuadra et al. (2015) have done a great deal to simulate the accretion flow from its origins.  In between the two regimes, we are able to place significant constraint on the structure of the accretion flow within the Bondi capture radius by linking observations to simulated 2D RRIOS accretion flows with MCMC fitting, self-consistently modelling the inflow and outflow regions simultaneously for the first time.  This lends necessary boundary conditions for those that seek to understand the flow at its inner and outer limits.

\begin{itemize}
\item The best-fitting temperature at the Bondi radius is 1.3e7 (1.24e7, 1.38e7) K and is consistent with an origin of shocked stellar wind material of velocity $\sim$ 1000 km s$^{-1}$.
\item The best-fitting electron density at the Bondi radius is 101.6 (91.4,111.1) cm$^{-3}$ and is consistent with estimates of stellar mass loss from stellar winds in the central cluster.
\item The angular momentum of captured gas, as parametrized by the centrifugal radius, is best fit as $r_\mathrm{c} = 0.058(\pm0.006) r_\mathrm{b}$.  This is the first observational constraint on the centrifugal radius, and provides an important condition for modelling as we move forward.
\item Low angular momentum accretion (Bondi-like) leads to too steep a density profile to spatially model the observed emission simultaneously at small and large radii.
\item We find the unresolved point-like quiescent emission is too steep ($\alpha$ = 4.8(3.5,7.5)) to be characterized by bremsstrahlung emission or undetected flaring emission of the same spectral shape as those of detected flares.  This emission is likely due to a combination of inverse-Compton scattering of low-frequency synchrotron emission by thermal electrons and synchrotron emission from a small percentage of electrons that are accelerated into a power-law tail.
\item The mass inflow rate at $r_\mathrm{b}$ is $\sim$ 10$^{-4}$ M$_\odot$ yr$^{-1}$.  This rate is well below the expected gas supply due to stellar wind mass loss in the vicinity of Sgr A*.
\item The radial profile of mass inflow is incredibly steep due to the strong balancing outflow, resulting in a mass inflow rate $\sim$ 10$^{-6}$ M$_\odot$ yr$^{-1}$ at $\sim10^3r_\mathrm{s}$, consistent with simulations of stellar wind dynamics.  This directly implies a net mass accretion rate of $\leq10^{-6}$ M$_\odot$ yr$^{-1}$ on to Sgr A*.  Extrapolating the mass inflow profile to lower radii results in an accretion rate at low radii that is consistent with estimates from Faraday rotation measures.
\item The polar outflow has an opening angle of 130-140$^\circ$ and a velocity of $\approx 350$ km s$^{-1}$.  We expect this predicted velocity to increase with inclusion of magnetic fields.  The effects of this polar outflow should be observable, either through its impact with the surrounding interstellar medium (ISM) or spatially resolved X-ray spectroscopic studies of the accretion flow.
\end{itemize}

The work herein is comprehensive, giving the first globally consistent picture of the Sgr A* accretion flow.  However, there is much that can be done to verify and push this study further as computational power grows and the next generation of X-ray telescopes are launched.  From the computational side, we can begin by running these simulations in 3D.  This will allow us to include magnetic fields, thereby self-consistently modelling the viscosity and feedback in the accretion flow.  The next step would be to realistically simulate the gas from its origin to within the centrifugal radius.  Observationally, a better determination of the truly quiescent IR flux is of paramount importance, which provides the strongest constraint on the non-thermal electron population.  In X-rays, with the release of \textit{Athena}, and ideally the \textit{X-ray Surveyor}, we may be able to more directly check our results by doing high-resolution spectroscopy with its non-dispersive spectrometer and far superior effective area ($\sim100$ times of \textit{Chandra}).  Leveraging such an instrument to extract the dynamics of individual lines will allow us to greatly constrain the flow structure, to the point of potentially mapping the inflow and outflow regions.

\section{Acknowledgements}
We would like to thank the anonymous referee for thoughtful comments.  We would like to thank Feng Yuan, Robert Weir, Andrew Degroot, Siming Liu, and Avery Broderick for their insightful and useful comments.  SRR and QDW are supported by the NASA/CXC grant number TM3-14006X.  Y-FJ is supported by NASA through Einstein Postdoctoral Fellowship grant number PF3-140109 awarded by the Chandra X-ray Center, which is operated by the Smithsonian Astrophysical Observatory for NASA under contract NAS8-03060.
\bibliographystyle{aa}
\bibliography{sgras_fitting}

\appendix
\section{Goodness of Fit}\label{app:goodness}

Looking again at Fig.~\ref{f:images}, we see that there are no apparent residuals within the inner $\sim$1 arcsec, however, beyond this, there may be significant residual effects.  It should be noted there are multiple contributions to the per pixel error that are not being modelled, including discrete X-ray sources, possible unmodelled extended structure, and time averaging of a chaotic simulation.  To exactly what extent these lead to an underestimate of the error, it is difficult to quantify. However, we will point out why, all things considered, we believe we have obtained a reasonable fit.

Let us first consider the time averaging of the simulation, despite the actual simulation being quite chaotic and turbulent as the outflowing gas collides with the inflowing gas.  Indeed, the temporal dispersion of a single grid cell can lead to a flux difference of a factor of 2 in either direction, probably leading to a slight underestimate of the size of our confidence intervals.  How much effect this would have after smoothing is unclear, since individual grid cells are not independent, but have clear turbulent structure between themselves during the simulation.  It is clear, though that at any particular point in time, the emission from a single point in space can deviate significantly.  However, these deviations are the strongest at low radii, which contribute relatively little to the overall flux in the image, and would likely be smoothed out by not only the PSF, but also the emission from the outer parts of the flow.

\begin{figure}
\centerline{\epsfig{figure=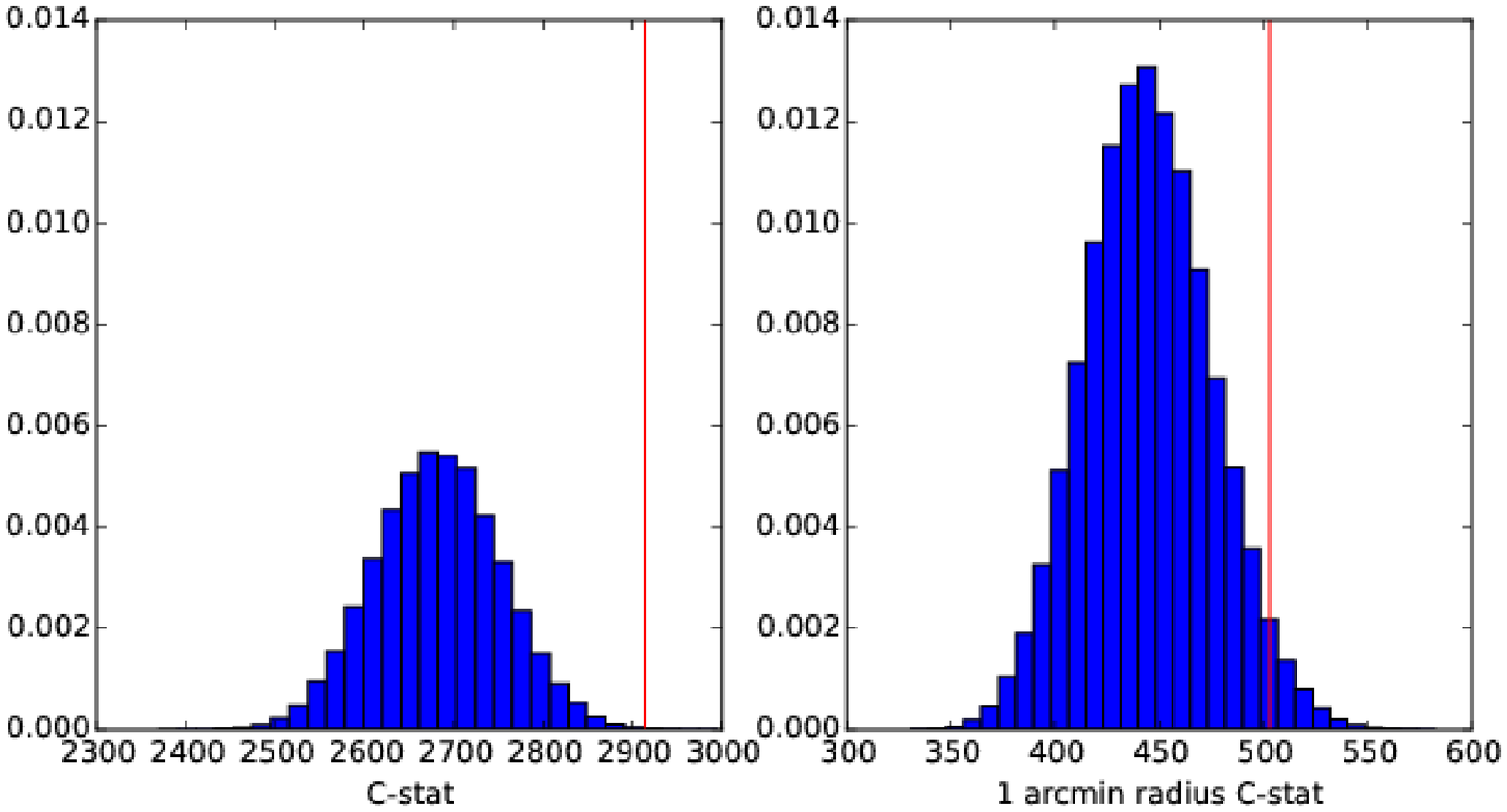,width=0.5\textwidth,angle=0}}
\caption{
Best-fitting C-statistic comparison to Poisson fluctuations for the $free$ model.  The red line indicates the best-fitting model compared to the data while the distributions are random realizations of the model.}
\label{f:goodness}
\end{figure}

Now, let us consider the effects that operate at larger radii.  One such source of uncertainty in the residuals is the number statistical fluctuations of discrete X-ray sources.  However, it is likely this term contributes negligibly to the emission and is merely absorbed into the background component.  More importantly, is unmodelled extended sub-structure that skirts the fitted region, such as dispersed emission and material from the G2 object, or the discreteness of the stellar winds and their colliding shocks. The apparent south-east excess, excluded from our fit, may represent an extreme case of such substructures. The presence of similar, probably fainter subsubstructure in the fitting region is echoed in the goodness of fit estimate.  As we can see in Fig.~\ref{f:goodness}, when excluding the pixels beyond 1 arcsec, the fit becomes remarkably better; the formal statistical consistency increases from ~0.1 to ~2.6\%.  Further, and importantly, by performing several fits masking beyond different radii, we find the parameter results are not sensitive to the fitted region within the image, leading us to conclude that the component does not contaminate our parameter estimates.

\end{CJK*}
\end{document}